\documentstyle[12pt]{article}
\def \be{\begin{equation}}
\def \ee{\end{equation}}
\def \bea{\begin{eqnarray}}
\def \eea{\end{eqnarray}}

\title{Monte Carlo Hamiltonian: the Linear Potentials\thanks{L.X.Q. 
is supported by the
National Natural Science Fund for Distinguished Young Scholars,
National Natural Science Foundation of China, 
Ministry of Education, 
Foundation of
the Zhongshan University Advanced Research Center
and Guangdong Provincial Natural Science Foundation.
H.K. is support by NSERC Canada.}
}
\vspace{2cm}
\author{Xiang-Qian Luo,
Jin-Jiang Liu, Chun-Qing Huang, Jun-Qin Jiang\\
{\small\sl Department of Physics, Zhongshan University, 
Guangzhou 510275, China}\\
Helmut Kr\"oger\\
{\small\sl D\'epartement de Physique, Universit\'e Laval, Qu\'ebec, 
Qu\'ebec G1K 7P4, Canada}\\
}

\date{\today}

\begin{document}
\maketitle

\begin{abstract}
We further study the validity of the Monte Carlo Hamiltonian method. 
The advantage of the method, in comparison with the standard Monte Carlo Lagrangian approach,
is its capability to study the excited states.
We consider two quantum mechanical models:
a symmetric one $V(x) = |x|/2$; and an asymmetric one
$V(x)=\infty$, for $x < 0$ and $V(x)=x$, for $x \ge  0$. 
The results for the spectrum, wave functions and thermodynamical observables
are in agreement with the analytical or Runge-Kutta calculations.
\end{abstract}

\vspace{1cm}


\vskip 1cm
\noindent
Published in {\it Communications in Theoretical Physics} {\bf 38} (2002) 561-565.

\setcounter{page}{0}
\newpage

\section{Introduction}
\label{sec:Introduction}

Path integral quantization
in the Lagrangian formulation and canonical quantization
in the Hamiltonian formulation are two ways
to quantize classical systems. 
Either one has advantages and disadvantages.
The Lagrangian formulation is suitable for numerical simulations on a computer
via Monte Carlo (MC). 
The enormous success of lattice gauge theory 
over the last two and half decades is certainly 
due to the fact that the MC method 
with importance sampling\cite{kn:Metropolis53}
is an excellent technique 
to compute high dimensional (and even ``infinite" dimensional) 
integrals.

Unfortunately, using the Lagrangian formulation 
it is difficult to estimate wave functions and 
the spectrum of excited states. 
Wave functions in conjunction with the energy spectrum 
contain more physical information than the energy spectrum alone. 
Although lattice QCD simulations in the Lagrangian formulation
give good estimates of the hadron masses,
one is yet far from a comprehensive understanding of
hadrons. Let us take as an example a new type of hadrons made of
gluons, the so-called glueballs. 
Lattice QCD calculations\cite{kn:Luo96}
predict the mass of the lightest glueball 
with quantum number $J^{PC}=0^{++}$,
to be $1650 \pm 100 MeV$. Experimentally, there are at least two
candidates: $f_0(1500)$ and $f_J(1710)$.
The investigation of the glueball production and decays  
can certainly provide additional important information for
experimental determination of a glueball. 
Therefore, it is important to be able to compute the glueball wave function.

In the Hamiltonian formulation,
one can obtain the ground state energy, 
but also wave functions and the spectrum of excited states.
Often, and in particular in the case of many-body systems, 
it is difficult to solve the stationary Schr\"odinger equation. 
Recently, we have suggested how to construct an effective Hamiltonian 
via MC\cite{kn:Jirari99} to describe the low energy spectrum 
and wave functions. 
The method has been tested 
in quantum mechanics\cite{kn:Jirari99,kn:MCHamilton} in 1,2 and 3 spatial dimensions (the free system,
the harmonic oscillator, the quartic potential, the local potential $V(x)=-V_0 \rm{sech}^2(x)$) and 
the scalar model  in quantum field theory\cite{scalar} (Klein-Gordon model and $\phi^4$ scalar model).

The purpose of this paper is to further test the viability of the MC Hamiltonian method 
in case of two potentials which generate a bound state spectrum:  
\bea
V(x) = |x|/2,
\label{potential1}  
\eea
and
\begin{eqnarray}
              V(x) =\bigg \{
                       \begin{array}{cc}
                         \infty,        & x < 0 \\
                         x,            &  x \ge  0. 
                  \end{array}
\label{potential2}
\end{eqnarray}
These two models have their own interest in quantum theory.
They have the form of a static quark-antiquark confining potential occurring in QCD.
The second potential describes the particle moving above the surface of the ground.  
In our previous investigations, only symmetric potentials were considered. 
Here we would like to see if the method works for linear potentials and those with infinity
and asymmetry.

The remainder of the paper is organized as follows. In Sect. 2,
the basic ideas of the effective Hamiltonian method is briefly discussed.
In Sect. 3, the numerical results are given. 
The conclusions are summarized in
Sect. 4.

\section{Effective Hamiltonian}
\label{sec:Effective Hamiltonian}

The construction of an effective Hamiltonian starting from a regular basis 
in position space has been proposed in Ref. \cite{kn:Jirari99}. 
Following Feynman's path integral formulation \cite{kn:Feynman65}, 
we consider the transition amplitude in imaginary time 
between time $t=0$ and $t=T$.  
Using imaginary time makes the path integral mathematically well defined, 
and renders it amenable to numerical simulations by MC.
Because the effective Hamiltonian is time independent, 
its construction in imaginary time should give the same result as in real time.
We consider the transition amplitude for all combinations of positions 
$x_{i}, x_{j} \in \{x_{1},\dots,x_{N}\}$,
\be
M_{i,j}(T)
= <x_{i} | e^{-H T/\hbar} | x_{j}>
= \int [dx] \exp[ -S_{E}[x]/\hbar ]\bigg |_{x_{j},0}^{x_{i},T} ,
\label{eq:DefHeff}               
\ee
where $S_{E}$ denotes the Euclidean action.
From the transition amplitudes $M_{ij}(T)$, 
one can construct the matrix
\be
M(T)=[M_{i,j}(T)]_{N \times N}.
\ee
$M(T)$ is a positive, Hermitian and symmetric matrix. 
$M$ can be factorized into a unitary matrix $U$ 
and a real diagonal matrix $D(T)$, such that
\be
\label{eq:DecompM}
M(T)=U^{\dagger}D(T)U.
\label{eq:MUD}
\ee
Then from $Eq.(\ref{eq:DefHeff})$, $Eq.(\ref{eq:MUD})$ one can identify
\be
\label{eq:EigenValVec}
U^{\dagger}_{i,k}=<x_i|E_k^{eff}>, ~~ D_k(T)=e^{-{E_k^{eff}}T/\hbar}.
\label{eq:UD}
\ee
The $k-th$ eigenvector $|E_k^{eff}>$ 
can be identified with the $k-th$ column of matrix $U^{\dagger}$.
The energy eigenvalues $E^{eff}_{k}$ are obtained from the logarithm 
of the diagonal matrix elements of $D(T)$.
This yields an effective Hamiltonian, 
\be
H_{eff} = \sum_{k =1}^{N} | E^{eff}_{k} > E^{eff}_{k} < E^{eff}_{k} |.
\label{Heff}
\ee
We compute the matrix element $M_{i,j}(T)$ directly from 
the action via MC with importance sampling. 
For details see Ref.\cite{kn:Jirari99}.

%\subsection{Basis in Hilbert Space}

To get the correct scale for the spectrum, the 
position state $\vert x_k \rangle$   
at initial time $t_i$ or final time $t_f$
(Bargman states or box states) 
should be properly normalized. We denote a normalized basis 
of Hilbert states as 
$\vert e_k \rangle$, $k=1, ..., N$. In position space, it can be
expressed as
\begin{eqnarray}
e_k(x) =\{
\begin{array}{cc}
     1/\sqrt{\Delta x_k},        & x \in [x_k,x_{k+1}] \\
     0,                         &  x \notin  [x_k,x_{k+1}] 
\end{array}
\end{eqnarray}
where $\Delta x_k=x_{k+1}-x_k$.
There is some arbitrariness in choosing  a basis for the
initial and final states.  
The simplest choice is a basis with $\Delta x_k ={\rm const.}$, 
which is called the ``regular basis''.

\section{Numerical results}
\label{sec:Numerical}

For the potential in Eq. (\ref{potential1}) an exact solution of the
ground state energy and wave function is analytically known \cite{kn:Analytical}. 
They are given by
\bea
E_0^{Exact} &=& \frac{|\xi_{0}|}{2} ({\hbar}^2/m)^{1/3} = 0.5094({\hbar}^2/m)^{1/3} ,
\nonumber \\
\psi_0(x) &=& \sqrt{|\xi|}\left[J_\frac{1}{3}\bigg(\frac{2}{3}|\xi|^{3/2}\bigg)
+J_{-\frac{1}{3}}\bigg(\frac{2}{3}|\xi|^{3/2}\bigg)\right] ,
\eea
where $\xi = \left(|x|-2E_0\right)\left(m/\hbar^2 \right)^{1/3}$, 
$\xi_{0} = \xi|_{x=0}$, $|\xi_{0}| = (3 z_{0}/2)^{2/3}$, $z_{0}$ 
is the solution to the equation 
\begin{eqnarray}
J_{2/3}(z)-J_{-2/3}(z)=0, 
\end{eqnarray}
and
$J_{1/3}$ and $J_{-1/3}$ are Airy functions.
We compare two methods: the MC 
Hamiltonian method and the Runge-Kutta algorithm with the node theorem.
The latter one means that we solve the stationary Schr\"odinger equation 
expressed as differential equation in position space
and search for the discrete eigenvalue $E_{n}$ corresponding to the number of nodes of the wave function.
Here $n$ denotes the quantum number, being related Eq. (\ref{Heff}) to $k$ by $n=k-1$.  
Because the transition matrix is symmetric, 
we implement in the code
\begin{eqnarray}
M_{i,j}&=&\left(M_{i,j}+M_{j,i}\right)/2,
\nonumber \\
M_{j,i} &=& M_{i,j},
\end{eqnarray}
when calculating the matrix elements 
using the MC Hamiltonian method.
There is an additional symmetry for the potential in Eq. (\ref{potential1}):  $V(-x)=V(x)$ 
so that the matrix elements should satisfy 
$M_{N+1-i,N+1-j}=M_{i,j}$. Therefore, 
in the code
\begin{eqnarray}
M_{i,j}&=&\left(M_{i,j}+M_{N+1-i,N+1-j}\right)/2,
\nonumber \\
M_{N+1-i,N+1-j} &=& M_{i,j}.
\end{eqnarray}
The use of these symmetries lead to reduction of the numerical errors.
We have computed the energy spectrum, wave functions as well 
as thermodynamical observables such as average energy, 
specific heat and the partition function. 
The results can be summarized as follows.
Fig. \ref{fig1} shows the ground state 
wave function. 
One observes good agreement between the MC Hamiltonian  result and the exact solution.
Fig. \ref{fig2} shows a similar comparison for the wave function of the first excited state. 
Tab. \ref{Tab1} shows the data for the spectrum.

\begin{table}
\begin{center}
\begin{tabular}{|c|c|c|c|c|}\hline
n  &  $E_n^{Exact}$  & $E_n^{R.K.}$ &  $E_n^{MC}$  & $\Delta E_n^{MC}/E_n^{R.K.}$ \\ \hline
0	& 0.509	& 0.509	&  0.504 $\pm$ 0.002 &  $1.0\%$\\	
1	&     	& 1.169	&   1.171 $\pm$ 0.002 &  $0.2\%$\\	
2	&     	& 1.624	&   1.622 $\pm$ 0.002 & $0.1\%$\\	
3	&     	& 2.044	&  2.047 $\pm$ 0.003 & $0.2\%$\\ 
4      &                 & 2.410      &   2.407 $\pm$ 0.004   &   $0.1\%$\\
5    &             & 2.760  & 2.764  $\pm$ 0.007 &  $0.1\%$\\
6    &             & 3.082  &3.075 $\pm$ 0.021   & $0.2\%$\\
7   &             & 3.393  & 3.399 $\pm$ 0.034  & $0.2\%$\\
8   &             & 3.686  & 3.669 $\pm$ 0.078  & $0.5\%$\\
9   &             & 3.972 & 3.982  $\pm$ 0.019  & $0.3\%$\\
10 &              &4.244   & 4.202 $\pm$ 0.105 & $1.0\%$\\
11 &              &4.511   &4.532  $\pm$ 0.041 & $0.5\%$\\
12 &              &4.768  &4.683  $\pm$ 0.069  & $1.8\%$\\
13 &              &5.020  &5.074  $\pm$ 0.142  & $1.0\%$\\
14 &             &5.264  &5.155   $\pm$ 0.153 &  $2.1\%$\\
15  &            &5.504  &5.673     $\pm$ 0.137  & $3.1\%$\\
16  &            &5.738    &5.708  $\pm$ 0.117  & $0.5\%$\\\hline	
\end{tabular}
\end{center}
\caption{\label{Tab1} Spectrum corresponding 
to the potential in Eq. (\ref{potential1}).
Here $E_n^{Exact}$, $E_n^{R.K.}$ and $E_n^{M.C.}$ 
denote respectively the data from the analytic solution,
Runge-Kutta algorithm and MC Hamiltonian.
The input parameters are
$m=1$, $\hbar=1$, $\Delta x=1$ and $N=21$. The  statistical an  relative errors of $E_n^{MC}$ are also shown.
 The error of the Runge-Kutta algorithm is less than the last digit and not shown.}
\end{table}

\begin{table}
\begin{center}
\begin{tabular}{|c|c|c|c|}\hline
n  &  $E_n^{Exact}$  & $E_n^{MC}$             & $\Delta E_n^{MC}/E_n^{Exact}$ \\ \hline
0	& 1.856	& 1.774 $\pm$ 0.086	& 4.4 $\%$\\	
1	& 3.244	& 3.182 $\pm$ 0.112	& 1.9 $\%$\\	
2	& 4.382	& 4.346 $\pm$ 0.125	& 0.8 $\%$\\	
3	& 5.383	& 5.300 $\pm$ 0.082	& 1.6 $\%$\\ \hline	
\end{tabular}
\end{center}
\caption{\label{Tab2} Spectrum corresponding to the potential in Eq. (\ref{potential2}).
The input parameters are
$m=1$, $\hbar=1$, $\Delta x=1$ and $N=20$.}
\end{table}

\begin{figure}[htb]
\begin{center}
% GNUPLOT: LaTeX picture
\setlength{\unitlength}{0.240900pt}
\ifx\plotpoint\undefined\newsavebox{\plotpoint}\fi
\sbox{\plotpoint}{\rule[-0.200pt]{0.400pt}{0.400pt}}%
\begin{picture}(1500,900)(0,0)
\font\gnuplot=cmr10 at 10pt
\gnuplot
\sbox{\plotpoint}{\rule[-0.200pt]{0.400pt}{0.400pt}}%
\put(181.0,123.0){\rule[-0.200pt]{4.818pt}{0.400pt}}
\put(161,123){\makebox(0,0)[r]{-0.2}}
\put(1419.0,123.0){\rule[-0.200pt]{4.818pt}{0.400pt}}
\put(181.0,232.0){\rule[-0.200pt]{4.818pt}{0.400pt}}
\put(161,232){\makebox(0,0)[r]{0}}
\put(1419.0,232.0){\rule[-0.200pt]{4.818pt}{0.400pt}}
\put(181.0,341.0){\rule[-0.200pt]{4.818pt}{0.400pt}}
\put(161,341){\makebox(0,0)[r]{0.2}}
\put(1419.0,341.0){\rule[-0.200pt]{4.818pt}{0.400pt}}
\put(181.0,450.0){\rule[-0.200pt]{4.818pt}{0.400pt}}
\put(161,450){\makebox(0,0)[r]{0.4}}
\put(1419.0,450.0){\rule[-0.200pt]{4.818pt}{0.400pt}}
\put(181.0,559.0){\rule[-0.200pt]{4.818pt}{0.400pt}}
\put(161,559){\makebox(0,0)[r]{0.6}}
\put(1419.0,559.0){\rule[-0.200pt]{4.818pt}{0.400pt}}
\put(181.0,668.0){\rule[-0.200pt]{4.818pt}{0.400pt}}
\put(161,668){\makebox(0,0)[r]{0.8}}
\put(1419.0,668.0){\rule[-0.200pt]{4.818pt}{0.400pt}}
\put(181.0,777.0){\rule[-0.200pt]{4.818pt}{0.400pt}}
\put(161,777){\makebox(0,0)[r]{1}}
\put(1419.0,777.0){\rule[-0.200pt]{4.818pt}{0.400pt}}
\put(181.0,123.0){\rule[-0.200pt]{0.400pt}{4.818pt}}
\put(181,82){\makebox(0,0){-10}}
\put(181.0,757.0){\rule[-0.200pt]{0.400pt}{4.818pt}}
\put(307.0,123.0){\rule[-0.200pt]{0.400pt}{4.818pt}}
\put(307,82){\makebox(0,0){-8}}
\put(307.0,757.0){\rule[-0.200pt]{0.400pt}{4.818pt}}
\put(433.0,123.0){\rule[-0.200pt]{0.400pt}{4.818pt}}
\put(433,82){\makebox(0,0){-6}}
\put(433.0,757.0){\rule[-0.200pt]{0.400pt}{4.818pt}}
\put(558.0,123.0){\rule[-0.200pt]{0.400pt}{4.818pt}}
\put(558,82){\makebox(0,0){-4}}
\put(558.0,757.0){\rule[-0.200pt]{0.400pt}{4.818pt}}
\put(684.0,123.0){\rule[-0.200pt]{0.400pt}{4.818pt}}
\put(684,82){\makebox(0,0){-2}}
\put(684.0,757.0){\rule[-0.200pt]{0.400pt}{4.818pt}}
\put(810.0,123.0){\rule[-0.200pt]{0.400pt}{4.818pt}}
\put(810,82){\makebox(0,0){0}}
\put(810.0,757.0){\rule[-0.200pt]{0.400pt}{4.818pt}}
\put(936.0,123.0){\rule[-0.200pt]{0.400pt}{4.818pt}}
\put(936,82){\makebox(0,0){2}}
\put(936.0,757.0){\rule[-0.200pt]{0.400pt}{4.818pt}}
\put(1062.0,123.0){\rule[-0.200pt]{0.400pt}{4.818pt}}
\put(1062,82){\makebox(0,0){4}}
\put(1062.0,757.0){\rule[-0.200pt]{0.400pt}{4.818pt}}
\put(1187.0,123.0){\rule[-0.200pt]{0.400pt}{4.818pt}}
\put(1187,82){\makebox(0,0){6}}
\put(1187.0,757.0){\rule[-0.200pt]{0.400pt}{4.818pt}}
\put(1313.0,123.0){\rule[-0.200pt]{0.400pt}{4.818pt}}
\put(1313,82){\makebox(0,0){8}}
\put(1313.0,757.0){\rule[-0.200pt]{0.400pt}{4.818pt}}
\put(1439.0,123.0){\rule[-0.200pt]{0.400pt}{4.818pt}}
\put(1439,82){\makebox(0,0){10}}
\put(1439.0,757.0){\rule[-0.200pt]{0.400pt}{4.818pt}}
\put(181.0,123.0){\rule[-0.200pt]{303.052pt}{0.400pt}}
\put(1439.0,123.0){\rule[-0.200pt]{0.400pt}{157.549pt}}
\put(181.0,777.0){\rule[-0.200pt]{303.052pt}{0.400pt}}
\put(40,450){\makebox(0,0){$\psi_0(x)$}}
\put(810,21){\makebox(0,0){x}}
\put(810,900){\makebox(0,0){$V(x)=x$ for $x \ge 0$, $V(x)=\infty$ for $x < 0$}} 
\put(810,839){\makebox(0,0){$m=1$, $T=1$, $\hbar=1$, $\Delta x=1$, $N=20$}}
\put(181.0,123.0){\rule[-0.200pt]{0.400pt}{157.549pt}}
\put(244,232){\raisebox{-.8pt}{\makebox(0,0){$\Diamond$}}}
\put(307,232){\raisebox{-.8pt}{\makebox(0,0){$\Diamond$}}}
\put(370,232){\raisebox{-.8pt}{\makebox(0,0){$\Diamond$}}}
\put(433,232){\raisebox{-.8pt}{\makebox(0,0){$\Diamond$}}}
\put(495,232){\raisebox{-.8pt}{\makebox(0,0){$\Diamond$}}}
\put(558,232){\raisebox{-.8pt}{\makebox(0,0){$\Diamond$}}}
\put(621,232){\raisebox{-.8pt}{\makebox(0,0){$\Diamond$}}}
\put(684,232){\raisebox{-.8pt}{\makebox(0,0){$\Diamond$}}}
\put(747,232){\raisebox{-.8pt}{\makebox(0,0){$\Diamond$}}}
\put(810,232){\raisebox{-.8pt}{\makebox(0,0){$\Diamond$}}}
\put(873,299){\raisebox{-.8pt}{\makebox(0,0){$\Diamond$}}}
\put(936,685){\raisebox{-.8pt}{\makebox(0,0){$\Diamond$}}}
\put(999,467){\raisebox{-.8pt}{\makebox(0,0){$\Diamond$}}}
\put(1062,283){\raisebox{-.8pt}{\makebox(0,0){$\Diamond$}}}
\put(1124,235){\raisebox{-.8pt}{\makebox(0,0){$\Diamond$}}}
\put(1187,232){\raisebox{-.8pt}{\makebox(0,0){$\Diamond$}}}
\put(1250,232){\raisebox{-.8pt}{\makebox(0,0){$\Diamond$}}}
\put(1313,232){\raisebox{-.8pt}{\makebox(0,0){$\Diamond$}}}
\put(1376,232){\raisebox{-.8pt}{\makebox(0,0){$\Diamond$}}}
\put(1439,232){\raisebox{-.8pt}{\makebox(0,0){$\Diamond$}}}
\put(244,232){\usebox{\plotpoint}}
\multiput(835.00,232.59)(1.033,0.482){9}{\rule{0.900pt}{0.116pt}}
\multiput(835.00,231.17)(10.132,6.000){2}{\rule{0.450pt}{0.400pt}}
\multiput(847.58,238.00)(0.493,0.734){23}{\rule{0.119pt}{0.685pt}}
\multiput(846.17,238.00)(13.000,17.579){2}{\rule{0.400pt}{0.342pt}}
\multiput(860.58,257.00)(0.492,1.573){21}{\rule{0.119pt}{1.333pt}}
\multiput(859.17,257.00)(12.000,34.233){2}{\rule{0.400pt}{0.667pt}}
\multiput(872.58,294.00)(0.492,2.564){21}{\rule{0.119pt}{2.100pt}}
\multiput(871.17,294.00)(12.000,55.641){2}{\rule{0.400pt}{1.050pt}}
\multiput(884.58,354.00)(0.492,3.340){21}{\rule{0.119pt}{2.700pt}}
\multiput(883.17,354.00)(12.000,72.396){2}{\rule{0.400pt}{1.350pt}}
\multiput(896.58,432.00)(0.492,3.857){21}{\rule{0.119pt}{3.100pt}}
\multiput(895.17,432.00)(12.000,83.566){2}{\rule{0.400pt}{1.550pt}}
\multiput(908.58,522.00)(0.492,3.770){21}{\rule{0.119pt}{3.033pt}}
\multiput(907.17,522.00)(12.000,81.704){2}{\rule{0.400pt}{1.517pt}}
\multiput(920.58,610.00)(0.492,2.736){21}{\rule{0.119pt}{2.233pt}}
\multiput(919.17,610.00)(12.000,59.365){2}{\rule{0.400pt}{1.117pt}}
\multiput(932.58,674.00)(0.492,0.755){21}{\rule{0.119pt}{0.700pt}}
\multiput(931.17,674.00)(12.000,16.547){2}{\rule{0.400pt}{0.350pt}}
\multiput(944.58,688.26)(0.492,-1.013){21}{\rule{0.119pt}{0.900pt}}
\multiput(943.17,690.13)(12.000,-22.132){2}{\rule{0.400pt}{0.450pt}}
\multiput(956.58,660.53)(0.492,-2.176){21}{\rule{0.119pt}{1.800pt}}
\multiput(955.17,664.26)(12.000,-47.264){2}{\rule{0.400pt}{0.900pt}}
\multiput(968.58,608.28)(0.492,-2.564){21}{\rule{0.119pt}{2.100pt}}
\multiput(967.17,612.64)(12.000,-55.641){2}{\rule{0.400pt}{1.050pt}}
\multiput(980.58,548.14)(0.492,-2.607){21}{\rule{0.119pt}{2.133pt}}
\multiput(979.17,552.57)(12.000,-56.572){2}{\rule{0.400pt}{1.067pt}}
\multiput(992.58,488.39)(0.492,-2.219){21}{\rule{0.119pt}{1.833pt}}
\multiput(991.17,492.19)(12.000,-48.195){2}{\rule{0.400pt}{0.917pt}}
\multiput(1004.58,437.77)(0.492,-1.789){21}{\rule{0.119pt}{1.500pt}}
\multiput(1003.17,440.89)(12.000,-38.887){2}{\rule{0.400pt}{0.750pt}}
\multiput(1016.58,396.99)(0.493,-1.408){23}{\rule{0.119pt}{1.208pt}}
\multiput(1015.17,399.49)(13.000,-33.493){2}{\rule{0.400pt}{0.604pt}}
\multiput(1029.58,360.88)(0.492,-1.444){21}{\rule{0.119pt}{1.233pt}}
\multiput(1028.17,363.44)(12.000,-31.440){2}{\rule{0.400pt}{0.617pt}}
\multiput(1041.58,327.30)(0.492,-1.315){21}{\rule{0.119pt}{1.133pt}}
\multiput(1040.17,329.65)(12.000,-28.648){2}{\rule{0.400pt}{0.567pt}}
\multiput(1053.58,297.26)(0.492,-1.013){21}{\rule{0.119pt}{0.900pt}}
\multiput(1052.17,299.13)(12.000,-22.132){2}{\rule{0.400pt}{0.450pt}}
\multiput(1065.58,274.51)(0.492,-0.625){21}{\rule{0.119pt}{0.600pt}}
\multiput(1064.17,275.75)(12.000,-13.755){2}{\rule{0.400pt}{0.300pt}}
\multiput(1077.00,260.92)(0.600,-0.491){17}{\rule{0.580pt}{0.118pt}}
\multiput(1077.00,261.17)(10.796,-10.000){2}{\rule{0.290pt}{0.400pt}}
\multiput(1089.00,250.93)(0.874,-0.485){11}{\rule{0.786pt}{0.117pt}}
\multiput(1089.00,251.17)(10.369,-7.000){2}{\rule{0.393pt}{0.400pt}}
\multiput(1101.00,243.93)(1.033,-0.482){9}{\rule{0.900pt}{0.116pt}}
\multiput(1101.00,244.17)(10.132,-6.000){2}{\rule{0.450pt}{0.400pt}}
\multiput(1113.00,237.94)(1.651,-0.468){5}{\rule{1.300pt}{0.113pt}}
\multiput(1113.00,238.17)(9.302,-4.000){2}{\rule{0.650pt}{0.400pt}}
\multiput(1125.00,233.95)(2.472,-0.447){3}{\rule{1.700pt}{0.108pt}}
\multiput(1125.00,234.17)(8.472,-3.000){2}{\rule{0.850pt}{0.400pt}}
\put(244.0,232.0){\rule[-0.200pt]{142.372pt}{0.400pt}}
\put(1137.0,232.0){\rule[-0.200pt]{72.752pt}{0.400pt}}
\end{picture}
\end{center}
\caption{Ground state wave function $\psi_0(x)$ corresponding to the potential in Eq. (\ref{potential2}),
where the continuous line represents the exact solution and the symbols stand for the data from
the MC Hamiltonian method.}
\label{fig3}
\end{figure}
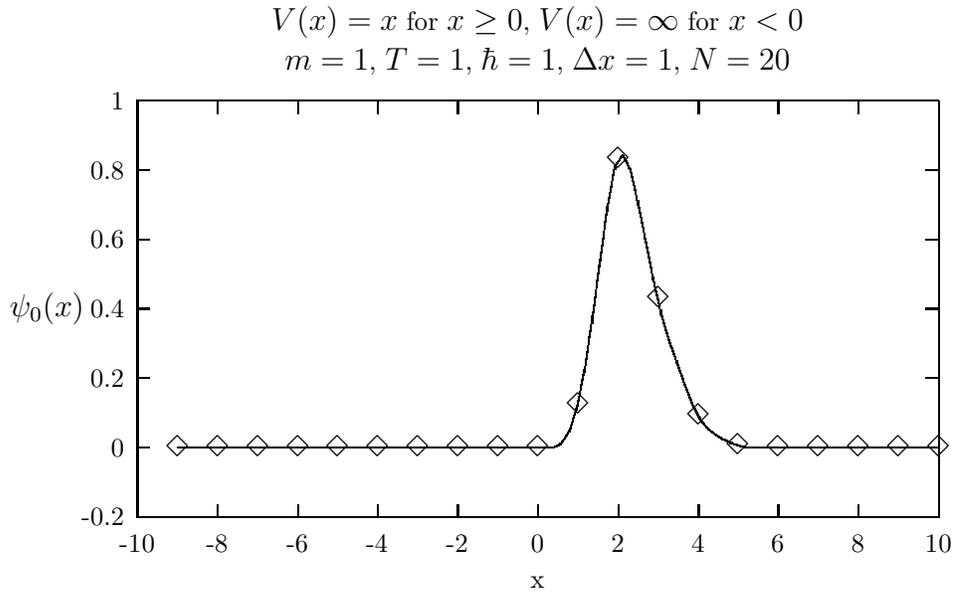

The model described by 
the potential in Eq. (\ref{potential2}) is exactly solvable.
The spectrum and wave function are given by
\bea
E_n &=& \lambda_{n} ({\hbar}^2/2m)^{1/3},
\nonumber \\
\psi_n(x) &=& 
\bigg \{
                       \begin{array}{cc}
                         0,        & x < 0 \\
\sqrt{|\xi_n|} \left[J_\frac{1}{3}\bigg(\frac{2}{3}|\xi|^{3/2}\bigg)
+J_{-\frac{1}{3}}\bigg(\frac{2}{3}|\xi|^{3/2}\bigg)\right] ,                          
                               &  x \ge  0, 
                  \end{array}
\label{exact2}
\eea
where $\xi_n =\left(x-E_n\right)\left(2m/\hbar^2 \right)^{1/3}$, and
$\lambda_n$ 
is the solution to the equation 
\begin{eqnarray}
J_\frac{1}{3}\bigg(\frac{2}{3}|\lambda_n|^{3/2}\bigg)
+J_{-\frac{1}{3}}\bigg(\frac{2}{3}|\lambda_n|^{3/2}\bigg)=0. 
\end{eqnarray}
Here 
are the first few: $\lambda_0=2.338$, $\lambda_1=4.088$, $\lambda_2=5.521$ and $\lambda_3=6.787$.
Fig. \ref{fig3}, Fig. \ref{fig4} and Tab.\ref{Tab2}
show a comparison of the spectrum and wave functions 
corresponding to the potential in Eq. (\ref{potential2}). 
Although the agreement is good,
the relative errors are bigger, in comparison with the first case.
This might be due to the lack of parity symmetry in the potential 
Eq. (\ref{potential2}).

The partition function is defined by $Z(\beta) = Tr(e^{-{\beta}H})$.
The average energy is defined by $U(\beta) = Tr(He^{-{\beta}H})/Z$,
and the specific heat by 
$C(\beta) = \partial U/ \partial \tau$.
We have used the following notation:
$\beta={(k_B \tau)}^{-1}$, $\tau$ is the temperature, $k_B$ is the Boltzmann constant, 
and we identify $\beta$ with the imaginary time $T$ 
by $\beta={T}/{\hbar}$. 
Since we have approximated $H$ by $H_{\rm{eff}}$, 
we can express those thermodynamical observables  
via the eigenvalues of the effective Hamiltonian
\begin{eqnarray}
Z^{\rm{eff}}(\beta) &=& \sum_{k=1}^{N}e^{-\beta E_{k}^{\rm{eff}}},
\nonumber \\
U^{\rm{eff}}(\beta) &=& \sum_{k=1}^{N}
{E_{k}^{\rm{eff}} {\rm e}^{-\beta E_{k}^{\rm{eff}}} \over Z^{\rm{eff}}(\beta)},
\nonumber \\
C^{\rm{eff}}(\beta) &=& k_B{\beta}^2\left(\sum_{k=1}^{N}
{(E_{k}^{\rm{eff}})^2e^{-\beta E_{k}^{\rm{eff}}} \over
Z^{\rm{eff}}(\beta)}-\left(U^{\rm{eff}}(\beta)\right)^2 \right).
\label{thermo}
\end{eqnarray}
The numerical results are shown in Fig.\ref{fig5} to Fig. \ref{fig8}.  
One observes good agreement between the results from the MC Hamiltonian 
and the reference solution.

\section{Conclusion}
\label{sec:Conclusion}

In the preceding sections, we have applied the MC Hamiltonian method\cite{kn:Jirari99}
to the linear potentials: one is symmetric, and another is asymmetric and has infinity. 
The results for the spectra, wave functions and
thermodynamical observables are in good agreement with other methods.
We also observe that the use of the parity symmetry whenever exists,
can lead to systematic reduction of numerical errors.

In most cases, where no exact solution is available,
numerical solution is required. Runge-Kutta method works only in 1D and
it is very difficult to obtain information on the excited states by standard Lagrangian MC approach.
In our opinion, the MC Hamiltonian has an advantage in these cases.

The MC Hamiltonian method approximates a quantum system with infinite states 
by a finite system with dimension $N$.
Increasing $N$, we can obtain more excited states, and reduce the systematic errors.
Larger $N$ value is also necessary when approaching the smaller $\beta$ (or higher temperature regime). However, 
the computation of the matrix elements is very time consuming.
In this paper, we used the regular basis, which is good enough for few body
quantum mechanics and might not be very efficient for
quantum field theory. In Ref. \cite{scalar}, we suggested the concept of
MC Hamiltonian using the 
stochastic basis and applied it successfully to the Klein-Gordon model and $\phi^4$ scalar model.
We believe that the application to  more complicated systems will be very interesting.

\newpage

\input fig1.tex

\input fig2.tex

\input fig4.tex

\input fig5.tex

\input fig6.tex

\begin{figure}[htb]
\begin{center}
% GNUPLOT: LaTeX picture
\setlength{\unitlength}{0.240900pt}
\ifx\plotpoint\undefined\newsavebox{\plotpoint}\fi
\sbox{\plotpoint}{\rule[-0.200pt]{0.400pt}{0.400pt}}%
\begin{picture}(1500,900)(0,0)
\font\gnuplot=cmr10 at 10pt
\gnuplot
\sbox{\plotpoint}{\rule[-0.200pt]{0.400pt}{0.400pt}}%
\put(161.0,123.0){\rule[-0.200pt]{4.818pt}{0.400pt}}
\put(141,123){\makebox(0,0)[r]{0}}
\put(1419.0,123.0){\rule[-0.200pt]{4.818pt}{0.400pt}}
\put(161.0,232.0){\rule[-0.200pt]{4.818pt}{0.400pt}}
\put(141,232){\makebox(0,0)[r]{0.5}}
\put(1419.0,232.0){\rule[-0.200pt]{4.818pt}{0.400pt}}
\put(161.0,341.0){\rule[-0.200pt]{4.818pt}{0.400pt}}
\put(141,341){\makebox(0,0)[r]{1}}
\put(1419.0,341.0){\rule[-0.200pt]{4.818pt}{0.400pt}}
\put(161.0,450.0){\rule[-0.200pt]{4.818pt}{0.400pt}}
\put(141,450){\makebox(0,0)[r]{1.5}}
\put(1419.0,450.0){\rule[-0.200pt]{4.818pt}{0.400pt}}
\put(161.0,559.0){\rule[-0.200pt]{4.818pt}{0.400pt}}
\put(141,559){\makebox(0,0)[r]{2}}
\put(1419.0,559.0){\rule[-0.200pt]{4.818pt}{0.400pt}}
\put(161.0,668.0){\rule[-0.200pt]{4.818pt}{0.400pt}}
\put(141,668){\makebox(0,0)[r]{2.5}}
\put(1419.0,668.0){\rule[-0.200pt]{4.818pt}{0.400pt}}
\put(161.0,777.0){\rule[-0.200pt]{4.818pt}{0.400pt}}
\put(141,777){\makebox(0,0)[r]{3}}
\put(1419.0,777.0){\rule[-0.200pt]{4.818pt}{0.400pt}}
\put(161.0,123.0){\rule[-0.200pt]{0.400pt}{4.818pt}}
\put(161,82){\makebox(0,0){0}}
\put(161.0,757.0){\rule[-0.200pt]{0.400pt}{4.818pt}}
\put(289.0,123.0){\rule[-0.200pt]{0.400pt}{4.818pt}}
\put(289,82){\makebox(0,0){2}}
\put(289.0,757.0){\rule[-0.200pt]{0.400pt}{4.818pt}}
\put(417.0,123.0){\rule[-0.200pt]{0.400pt}{4.818pt}}
\put(417,82){\makebox(0,0){4}}
\put(417.0,757.0){\rule[-0.200pt]{0.400pt}{4.818pt}}
\put(544.0,123.0){\rule[-0.200pt]{0.400pt}{4.818pt}}
\put(544,82){\makebox(0,0){6}}
\put(544.0,757.0){\rule[-0.200pt]{0.400pt}{4.818pt}}
\put(672.0,123.0){\rule[-0.200pt]{0.400pt}{4.818pt}}
\put(672,82){\makebox(0,0){8}}
\put(672.0,757.0){\rule[-0.200pt]{0.400pt}{4.818pt}}
\put(800.0,123.0){\rule[-0.200pt]{0.400pt}{4.818pt}}
\put(800,82){\makebox(0,0){10}}
\put(800.0,757.0){\rule[-0.200pt]{0.400pt}{4.818pt}}
\put(928.0,123.0){\rule[-0.200pt]{0.400pt}{4.818pt}}
\put(928,82){\makebox(0,0){12}}
\put(928.0,757.0){\rule[-0.200pt]{0.400pt}{4.818pt}}
\put(1056.0,123.0){\rule[-0.200pt]{0.400pt}{4.818pt}}
\put(1056,82){\makebox(0,0){14}}
\put(1056.0,757.0){\rule[-0.200pt]{0.400pt}{4.818pt}}
\put(1183.0,123.0){\rule[-0.200pt]{0.400pt}{4.818pt}}
\put(1183,82){\makebox(0,0){16}}
\put(1183.0,757.0){\rule[-0.200pt]{0.400pt}{4.818pt}}
\put(1311.0,123.0){\rule[-0.200pt]{0.400pt}{4.818pt}}
\put(1311,82){\makebox(0,0){18}}
\put(1311.0,757.0){\rule[-0.200pt]{0.400pt}{4.818pt}}
\put(1439.0,123.0){\rule[-0.200pt]{0.400pt}{4.818pt}}
\put(1439,82){\makebox(0,0){20}}
\put(1439.0,757.0){\rule[-0.200pt]{0.400pt}{4.818pt}}
\put(161.0,123.0){\rule[-0.200pt]{307.870pt}{0.400pt}}
\put(1439.0,123.0){\rule[-0.200pt]{0.400pt}{157.549pt}}
\put(161.0,777.0){\rule[-0.200pt]{307.870pt}{0.400pt}}
\put(40,450){\makebox(0,0){$U$}}
\put(800,21){\makebox(0,0){$\beta$}}
\put(810,900){\makebox(0,0){$V(x)=x$ for $x \ge 0$, $V(x)=\infty$ for $x < 0$}} 
\put(810,839){\makebox(0,0){$m=1$, $T=1$, $\hbar=1$, $\Delta x=1$, $N=20$}}
\put(161.0,123.0){\rule[-0.200pt]{0.400pt}{157.549pt}}
\put(225,628){\raisebox{-.8pt}{\makebox(0,0){$\Diamond$}}}
\put(289,531){\raisebox{-.8pt}{\makebox(0,0){$\Diamond$}}}
\put(353,514){\raisebox{-.8pt}{\makebox(0,0){$\Diamond$}}}
\put(417,511){\raisebox{-.8pt}{\makebox(0,0){$\Diamond$}}}
\put(480,510){\raisebox{-.8pt}{\makebox(0,0){$\Diamond$}}}
\put(544,510){\raisebox{-.8pt}{\makebox(0,0){$\Diamond$}}}
\put(608,510){\raisebox{-.8pt}{\makebox(0,0){$\Diamond$}}}
\put(672,510){\raisebox{-.8pt}{\makebox(0,0){$\Diamond$}}}
\put(736,510){\raisebox{-.8pt}{\makebox(0,0){$\Diamond$}}}
\put(800,510){\raisebox{-.8pt}{\makebox(0,0){$\Diamond$}}}
\put(864,510){\raisebox{-.8pt}{\makebox(0,0){$\Diamond$}}}
\put(928,510){\raisebox{-.8pt}{\makebox(0,0){$\Diamond$}}}
\put(992,510){\raisebox{-.8pt}{\makebox(0,0){$\Diamond$}}}
\put(1056,510){\raisebox{-.8pt}{\makebox(0,0){$\Diamond$}}}
\put(1119,510){\raisebox{-.8pt}{\makebox(0,0){$\Diamond$}}}
\put(1183,510){\raisebox{-.8pt}{\makebox(0,0){$\Diamond$}}}
\put(1247,510){\raisebox{-.8pt}{\makebox(0,0){$\Diamond$}}}
\put(1311,510){\raisebox{-.8pt}{\makebox(0,0){$\Diamond$}}}
\put(1375,510){\raisebox{-.8pt}{\makebox(0,0){$\Diamond$}}}
\put(1439,510){\raisebox{-.8pt}{\makebox(0,0){$\Diamond$}}}
\put(225,647){\usebox{\plotpoint}}
\multiput(225.58,643.54)(0.492,-0.927){21}{\rule{0.119pt}{0.833pt}}
\multiput(224.17,645.27)(12.000,-20.270){2}{\rule{0.400pt}{0.417pt}}
\multiput(237.58,621.54)(0.492,-0.927){21}{\rule{0.119pt}{0.833pt}}
\multiput(236.17,623.27)(12.000,-20.270){2}{\rule{0.400pt}{0.417pt}}
\multiput(249.58,600.03)(0.493,-0.774){23}{\rule{0.119pt}{0.715pt}}
\multiput(248.17,601.52)(13.000,-18.515){2}{\rule{0.400pt}{0.358pt}}
\multiput(262.58,580.09)(0.492,-0.755){21}{\rule{0.119pt}{0.700pt}}
\multiput(261.17,581.55)(12.000,-16.547){2}{\rule{0.400pt}{0.350pt}}
\multiput(274.58,562.79)(0.492,-0.539){21}{\rule{0.119pt}{0.533pt}}
\multiput(273.17,563.89)(12.000,-11.893){2}{\rule{0.400pt}{0.267pt}}
\multiput(286.00,550.92)(0.600,-0.491){17}{\rule{0.580pt}{0.118pt}}
\multiput(286.00,551.17)(10.796,-10.000){2}{\rule{0.290pt}{0.400pt}}
\multiput(298.00,540.93)(1.123,-0.482){9}{\rule{0.967pt}{0.116pt}}
\multiput(298.00,541.17)(10.994,-6.000){2}{\rule{0.483pt}{0.400pt}}
\put(311,534.17){\rule{2.500pt}{0.400pt}}
\multiput(311.00,535.17)(6.811,-2.000){2}{\rule{1.250pt}{0.400pt}}
\put(323,532.67){\rule{2.891pt}{0.400pt}}
\multiput(323.00,533.17)(6.000,-1.000){2}{\rule{1.445pt}{0.400pt}}
\put(335,531.67){\rule{3.132pt}{0.400pt}}
\multiput(335.00,532.17)(6.500,-1.000){2}{\rule{1.566pt}{0.400pt}}
\put(372,530.67){\rule{2.891pt}{0.400pt}}
\multiput(372.00,531.17)(6.000,-1.000){2}{\rule{1.445pt}{0.400pt}}
\put(384,529.67){\rule{3.132pt}{0.400pt}}
\multiput(384.00,530.17)(6.500,-1.000){2}{\rule{1.566pt}{0.400pt}}
\put(397,528.67){\rule{2.891pt}{0.400pt}}
\multiput(397.00,529.17)(6.000,-1.000){2}{\rule{1.445pt}{0.400pt}}
\put(348.0,532.0){\rule[-0.200pt]{5.782pt}{0.400pt}}
\put(421,527.67){\rule{2.891pt}{0.400pt}}
\multiput(421.00,528.17)(6.000,-1.000){2}{\rule{1.445pt}{0.400pt}}
\put(409.0,529.0){\rule[-0.200pt]{2.891pt}{0.400pt}}
\put(433.0,528.0){\rule[-0.200pt]{242.345pt}{0.400pt}}
\end{picture}
\end{center}
\caption{Average energy $U({\beta})$ corresponding to the potential in Eq. (\ref{potential2}),
where the continuous line represents the result obtained by substituting the eigenvalues of 
the anlytic solution into Eq. (\ref{thermo}) and  the symbols stand for the data from
the MC Hamiltonian method.}
\label{fig7}
\end{figure}

\begin{figure}[htb]
\begin{center}
% GNUPLOT: LaTeX picture
\setlength{\unitlength}{0.240900pt}
\ifx\plotpoint\undefined\newsavebox{\plotpoint}\fi
\sbox{\plotpoint}{\rule[-0.200pt]{0.400pt}{0.400pt}}%
\begin{picture}(1500,900)(0,0)
\font\gnuplot=cmr10 at 10pt
\gnuplot
\sbox{\plotpoint}{\rule[-0.200pt]{0.400pt}{0.400pt}}%
\put(161.0,123.0){\rule[-0.200pt]{4.818pt}{0.400pt}}
\put(141,123){\makebox(0,0)[r]{0}}
\put(1419.0,123.0){\rule[-0.200pt]{4.818pt}{0.400pt}}
\put(161.0,232.0){\rule[-0.200pt]{4.818pt}{0.400pt}}
\put(141,232){\makebox(0,0)[r]{0.2}}
\put(1419.0,232.0){\rule[-0.200pt]{4.818pt}{0.400pt}}
\put(161.0,341.0){\rule[-0.200pt]{4.818pt}{0.400pt}}
\put(141,341){\makebox(0,0)[r]{0.4}}
\put(1419.0,341.0){\rule[-0.200pt]{4.818pt}{0.400pt}}
\put(161.0,450.0){\rule[-0.200pt]{4.818pt}{0.400pt}}
\put(141,450){\makebox(0,0)[r]{0.6}}
\put(1419.0,450.0){\rule[-0.200pt]{4.818pt}{0.400pt}}
\put(161.0,559.0){\rule[-0.200pt]{4.818pt}{0.400pt}}
\put(141,559){\makebox(0,0)[r]{0.8}}
\put(1419.0,559.0){\rule[-0.200pt]{4.818pt}{0.400pt}}
\put(161.0,668.0){\rule[-0.200pt]{4.818pt}{0.400pt}}
\put(141,668){\makebox(0,0)[r]{1}}
\put(1419.0,668.0){\rule[-0.200pt]{4.818pt}{0.400pt}}
\put(161.0,777.0){\rule[-0.200pt]{4.818pt}{0.400pt}}
\put(141,777){\makebox(0,0)[r]{1.2}}
\put(1419.0,777.0){\rule[-0.200pt]{4.818pt}{0.400pt}}
\put(161.0,123.0){\rule[-0.200pt]{0.400pt}{4.818pt}}
\put(161,82){\makebox(0,0){0}}
\put(161.0,757.0){\rule[-0.200pt]{0.400pt}{4.818pt}}
\put(289.0,123.0){\rule[-0.200pt]{0.400pt}{4.818pt}}
\put(289,82){\makebox(0,0){2}}
\put(289.0,757.0){\rule[-0.200pt]{0.400pt}{4.818pt}}
\put(417.0,123.0){\rule[-0.200pt]{0.400pt}{4.818pt}}
\put(417,82){\makebox(0,0){4}}
\put(417.0,757.0){\rule[-0.200pt]{0.400pt}{4.818pt}}
\put(544.0,123.0){\rule[-0.200pt]{0.400pt}{4.818pt}}
\put(544,82){\makebox(0,0){6}}
\put(544.0,757.0){\rule[-0.200pt]{0.400pt}{4.818pt}}
\put(672.0,123.0){\rule[-0.200pt]{0.400pt}{4.818pt}}
\put(672,82){\makebox(0,0){8}}
\put(672.0,757.0){\rule[-0.200pt]{0.400pt}{4.818pt}}
\put(800.0,123.0){\rule[-0.200pt]{0.400pt}{4.818pt}}
\put(800,82){\makebox(0,0){10}}
\put(800.0,757.0){\rule[-0.200pt]{0.400pt}{4.818pt}}
\put(928.0,123.0){\rule[-0.200pt]{0.400pt}{4.818pt}}
\put(928,82){\makebox(0,0){12}}
\put(928.0,757.0){\rule[-0.200pt]{0.400pt}{4.818pt}}
\put(1056.0,123.0){\rule[-0.200pt]{0.400pt}{4.818pt}}
\put(1056,82){\makebox(0,0){14}}
\put(1056.0,757.0){\rule[-0.200pt]{0.400pt}{4.818pt}}
\put(1183.0,123.0){\rule[-0.200pt]{0.400pt}{4.818pt}}
\put(1183,82){\makebox(0,0){16}}
\put(1183.0,757.0){\rule[-0.200pt]{0.400pt}{4.818pt}}
\put(1311.0,123.0){\rule[-0.200pt]{0.400pt}{4.818pt}}
\put(1311,82){\makebox(0,0){18}}
\put(1311.0,757.0){\rule[-0.200pt]{0.400pt}{4.818pt}}
\put(1439.0,123.0){\rule[-0.200pt]{0.400pt}{4.818pt}}
\put(1439,82){\makebox(0,0){20}}
\put(1439.0,757.0){\rule[-0.200pt]{0.400pt}{4.818pt}}
\put(161.0,123.0){\rule[-0.200pt]{307.870pt}{0.400pt}}
\put(1439.0,123.0){\rule[-0.200pt]{0.400pt}{157.549pt}}
\put(161.0,777.0){\rule[-0.200pt]{307.870pt}{0.400pt}}
\put(20,450){\makebox(0,0){$C/k_B$}}
\put(800,21){\makebox(0,0){$\beta$}}
\put(810,900){\makebox(0,0){$V(x)=x$ for $x \ge 0$, $V(x)=\infty$ for $x < 0$}} 
\put(810,839){\makebox(0,0){$m=1$, $T=1$, $\hbar=1$, $\Delta x=1$, $N=20$}}
\put(161.0,123.0){\rule[-0.200pt]{0.400pt}{157.549pt}}
\put(225,718){\raisebox{-.8pt}{\makebox(0,0){$\Diamond$}}}
\put(289,453){\raisebox{-.8pt}{\makebox(0,0){$\Diamond$}}}
\put(353,277){\raisebox{-.8pt}{\makebox(0,0){$\Diamond$}}}
\put(417,187){\raisebox{-.8pt}{\makebox(0,0){$\Diamond$}}}
\put(480,147){\raisebox{-.8pt}{\makebox(0,0){$\Diamond$}}}
\put(544,131){\raisebox{-.8pt}{\makebox(0,0){$\Diamond$}}}
\put(608,126){\raisebox{-.8pt}{\makebox(0,0){$\Diamond$}}}
\put(672,124){\raisebox{-.8pt}{\makebox(0,0){$\Diamond$}}}
\put(736,123){\raisebox{-.8pt}{\makebox(0,0){$\Diamond$}}}
\put(800,123){\raisebox{-.8pt}{\makebox(0,0){$\Diamond$}}}
\put(864,123){\raisebox{-.8pt}{\makebox(0,0){$\Diamond$}}}
\put(928,123){\raisebox{-.8pt}{\makebox(0,0){$\Diamond$}}}
\put(992,123){\raisebox{-.8pt}{\makebox(0,0){$\Diamond$}}}
\put(1056,123){\raisebox{-.8pt}{\makebox(0,0){$\Diamond$}}}
\put(1119,123){\raisebox{-.8pt}{\makebox(0,0){$\Diamond$}}}
\put(1183,123){\raisebox{-.8pt}{\makebox(0,0){$\Diamond$}}}
\put(1247,123){\raisebox{-.8pt}{\makebox(0,0){$\Diamond$}}}
\put(1311,123){\raisebox{-.8pt}{\makebox(0,0){$\Diamond$}}}
\put(1375,123){\raisebox{-.8pt}{\makebox(0,0){$\Diamond$}}}
\put(1439,123){\raisebox{-.8pt}{\makebox(0,0){$\Diamond$}}}
\put(225,720){\usebox{\plotpoint}}
\multiput(225.58,712.11)(0.492,-2.306){21}{\rule{0.119pt}{1.900pt}}
\multiput(224.17,716.06)(12.000,-50.056){2}{\rule{0.400pt}{0.950pt}}
\multiput(237.58,658.39)(0.492,-2.219){21}{\rule{0.119pt}{1.833pt}}
\multiput(236.17,662.19)(12.000,-48.195){2}{\rule{0.400pt}{0.917pt}}
\multiput(249.58,606.94)(0.493,-2.043){23}{\rule{0.119pt}{1.700pt}}
\multiput(248.17,610.47)(13.000,-48.472){2}{\rule{0.400pt}{0.850pt}}
\multiput(262.58,554.94)(0.492,-2.047){21}{\rule{0.119pt}{1.700pt}}
\multiput(261.17,558.47)(12.000,-44.472){2}{\rule{0.400pt}{0.850pt}}
\multiput(274.58,507.22)(0.492,-1.961){21}{\rule{0.119pt}{1.633pt}}
\multiput(273.17,510.61)(12.000,-42.610){2}{\rule{0.400pt}{0.817pt}}
\multiput(286.58,461.63)(0.492,-1.832){21}{\rule{0.119pt}{1.533pt}}
\multiput(285.17,464.82)(12.000,-39.817){2}{\rule{0.400pt}{0.767pt}}
\multiput(298.58,419.73)(0.493,-1.488){23}{\rule{0.119pt}{1.269pt}}
\multiput(297.17,422.37)(13.000,-35.366){2}{\rule{0.400pt}{0.635pt}}
\multiput(311.58,381.74)(0.492,-1.487){21}{\rule{0.119pt}{1.267pt}}
\multiput(310.17,384.37)(12.000,-32.371){2}{\rule{0.400pt}{0.633pt}}
\multiput(323.58,347.30)(0.492,-1.315){21}{\rule{0.119pt}{1.133pt}}
\multiput(322.17,349.65)(12.000,-28.648){2}{\rule{0.400pt}{0.567pt}}
\multiput(335.58,317.01)(0.493,-1.091){23}{\rule{0.119pt}{0.962pt}}
\multiput(334.17,319.00)(13.000,-26.004){2}{\rule{0.400pt}{0.481pt}}
\multiput(348.58,289.13)(0.492,-1.056){21}{\rule{0.119pt}{0.933pt}}
\multiput(347.17,291.06)(12.000,-23.063){2}{\rule{0.400pt}{0.467pt}}
\multiput(360.58,264.68)(0.492,-0.884){21}{\rule{0.119pt}{0.800pt}}
\multiput(359.17,266.34)(12.000,-19.340){2}{\rule{0.400pt}{0.400pt}}
\multiput(372.58,243.96)(0.492,-0.798){21}{\rule{0.119pt}{0.733pt}}
\multiput(371.17,245.48)(12.000,-17.478){2}{\rule{0.400pt}{0.367pt}}
\multiput(384.58,225.54)(0.493,-0.616){23}{\rule{0.119pt}{0.592pt}}
\multiput(383.17,226.77)(13.000,-14.771){2}{\rule{0.400pt}{0.296pt}}
\multiput(397.58,209.65)(0.492,-0.582){21}{\rule{0.119pt}{0.567pt}}
\multiput(396.17,210.82)(12.000,-12.824){2}{\rule{0.400pt}{0.283pt}}
\multiput(409.00,196.92)(0.496,-0.492){21}{\rule{0.500pt}{0.119pt}}
\multiput(409.00,197.17)(10.962,-12.000){2}{\rule{0.250pt}{0.400pt}}
\multiput(421.00,184.92)(0.543,-0.492){19}{\rule{0.536pt}{0.118pt}}
\multiput(421.00,185.17)(10.887,-11.000){2}{\rule{0.268pt}{0.400pt}}
\multiput(433.00,173.93)(0.824,-0.488){13}{\rule{0.750pt}{0.117pt}}
\multiput(433.00,174.17)(11.443,-8.000){2}{\rule{0.375pt}{0.400pt}}
\multiput(446.00,165.93)(0.758,-0.488){13}{\rule{0.700pt}{0.117pt}}
\multiput(446.00,166.17)(10.547,-8.000){2}{\rule{0.350pt}{0.400pt}}
\multiput(458.00,157.93)(1.033,-0.482){9}{\rule{0.900pt}{0.116pt}}
\multiput(458.00,158.17)(10.132,-6.000){2}{\rule{0.450pt}{0.400pt}}
\multiput(470.00,151.93)(1.267,-0.477){7}{\rule{1.060pt}{0.115pt}}
\multiput(470.00,152.17)(9.800,-5.000){2}{\rule{0.530pt}{0.400pt}}
\multiput(482.00,146.93)(1.378,-0.477){7}{\rule{1.140pt}{0.115pt}}
\multiput(482.00,147.17)(10.634,-5.000){2}{\rule{0.570pt}{0.400pt}}
\multiput(495.00,141.95)(2.472,-0.447){3}{\rule{1.700pt}{0.108pt}}
\multiput(495.00,142.17)(8.472,-3.000){2}{\rule{0.850pt}{0.400pt}}
\multiput(507.00,138.95)(2.472,-0.447){3}{\rule{1.700pt}{0.108pt}}
\multiput(507.00,139.17)(8.472,-3.000){2}{\rule{0.850pt}{0.400pt}}
\multiput(519.00,135.95)(2.472,-0.447){3}{\rule{1.700pt}{0.108pt}}
\multiput(519.00,136.17)(8.472,-3.000){2}{\rule{0.850pt}{0.400pt}}
\put(531,132.17){\rule{2.700pt}{0.400pt}}
\multiput(531.00,133.17)(7.396,-2.000){2}{\rule{1.350pt}{0.400pt}}
\put(544,130.67){\rule{2.891pt}{0.400pt}}
\multiput(544.00,131.17)(6.000,-1.000){2}{\rule{1.445pt}{0.400pt}}
\put(556,129.17){\rule{2.500pt}{0.400pt}}
\multiput(556.00,130.17)(6.811,-2.000){2}{\rule{1.250pt}{0.400pt}}
\put(568,127.67){\rule{3.132pt}{0.400pt}}
\multiput(568.00,128.17)(6.500,-1.000){2}{\rule{1.566pt}{0.400pt}}
\put(581,126.67){\rule{2.891pt}{0.400pt}}
\multiput(581.00,127.17)(6.000,-1.000){2}{\rule{1.445pt}{0.400pt}}
\put(593,125.67){\rule{2.891pt}{0.400pt}}
\multiput(593.00,126.17)(6.000,-1.000){2}{\rule{1.445pt}{0.400pt}}
\put(617,124.67){\rule{3.132pt}{0.400pt}}
\multiput(617.00,125.17)(6.500,-1.000){2}{\rule{1.566pt}{0.400pt}}
\put(605.0,126.0){\rule[-0.200pt]{2.891pt}{0.400pt}}
\put(642,123.67){\rule{2.891pt}{0.400pt}}
\multiput(642.00,124.17)(6.000,-1.000){2}{\rule{1.445pt}{0.400pt}}
\put(630.0,125.0){\rule[-0.200pt]{2.891pt}{0.400pt}}
\put(703,122.67){\rule{2.891pt}{0.400pt}}
\multiput(703.00,123.17)(6.000,-1.000){2}{\rule{1.445pt}{0.400pt}}
\put(654.0,124.0){\rule[-0.200pt]{11.804pt}{0.400pt}}
\put(715.0,123.0){\rule[-0.200pt]{174.412pt}{0.400pt}}
\end{picture}
\end{center}
\caption{Same as Fig. \ref{fig7}, but for specific heat $C({\beta})$.}
\label{fig8}
\end{figure}

\end{document}